%Paper: hep-th/9311146
%From: a.mikovic@ic.ac.uk
%Date: Wed, 24 Nov 93 15:26:00 gmt

%%%%%%%%%%%%%%%%%%%%%%%%%%%% MACROS %%%%%%%%%%%%%%%%%%%%%%%%%%%%%%%

%%%%%%%%%% BOXX %%%%%%%%%%

\def\sqr#1#2{{\vcenter{\hrule height.#2pt
             \hbox{\vrule width.#2pt height#1pt \kern#1pt
             \vrule width.#2pt}
             \hrule height.#2pt}}}

%%%%%%%%%% FONTS %%%%%%%%%%

\def\pmb#1{\setbox0=\hbox{#1}\kern-.025em
    \copy0\kern-\wd0\kern.05em\kern-.025em\raise.029em\box0}

\def\a{\alpha}         
\def\b{\beta}       \def\m{\mu}
          
\def\d{\delta}          \def\p{\pi}     
\def\e{\epsilon}                      
\def\ve{\varepsilon}                    
                         \def\t{\tau}
        \def\vphi{\varphi}

      \def\cO{{\cal O}}

%%%%%%%%%% DEF %%%%%%%%%%

\def\fr#1 #2{\hbox{${#1\over #2}$}}        %  {\fr #1 #2}
\def\leaderfill{\leaders\hbox to 1em{\hss.\hss}\hfill}

\def\section#1{
  \vskip.7cm\goodbreak
  \noindent{\bf #1}
  \nobreak\vskip.4cm\nobreak  }

\def\subsection#1{
  \vskip.3cm\goodbreak
  \noindent{\it #1}
  \vskip.3cm\nobreak}

\def\subsub#1{\par\vskip.3cm {\bf #1} }

\def\ver#1{\left\vert\vbox to #1mm{}\right.}

            % for References

%%%%%%% TITLE PAGE %%%%%%%%

\font\eightrm=cmr8                                % for footnote
\font\twelvebf=cmbx12                             % for title

\def\preprint#1{\hskip10cm #1 \par}
\def\date#1{\hskip10cm #1 \vskip2cm}
\def\title#1{ \centerline{\twelvebf\uppercase{#1}} }
\def\titlef#1{ \vskip.2cm                          % title with footnote
      \centerline{ \twelvebf\uppercase{#1}
                   \hskip-5pt ${\phantom{\ver{3}}}^\star$  }
      \vfootnote{$^\star$}{\eightrm Work supported in part
      by the Serbian Reserch Foundation, Yugoslavia.} \vskip.5cm }
\def\author#1{ \vskip.5cm \centerline{#1} }
\def\institution#1{ \centerline{\it #1} }
\def\abstract#1{ \vskip3cm \centerline{\bf Abstract}
                 \vskip.3cm {#1} \vfill\eject }

%%%%%%%%%%%%%%%%%%%%%%%%%%%%%%%%%%%%%%%%%%%%%%%%%%%%%%%%%%%%%%%%%%%%%%%%

\magnification=\magstep1
\null

\preprint{IF--S11/93}
\date{November, 1993.}

\title{Conformal gauge generators in}
\titlef{Liouville theory}

\author{M. Blagojevi\'c}
\institution{Institute of Physics, P.O.Box 57, 11001 Beograd, Yugoslavia}
\author{M. Vasili\'c and T. Vuka\v sinac}
\institution{Department of Theoretical Pysics, Institute of Nuclear
             Sciences, P.O.Box 522,}
\institution{11001 Beograd, Yugoslavia}

\abstract{The conformal symmetry in the Liouville theory is analysed by
using the Hamiltonian light--front formalism. The boundary conditions
of dynamical variables are seen to involve an arbitrary function of
time, so that the standard methods for studying gauge symmetries do not
work. We develop a general method for constructing the gauge
generators, which enables a consistent treatment of the boundary
conditions present in the case of the conformal symmetry.}

\section{1. Introduction} %%%%%%%%%%%%%%%%%%%%%%%%%%%%%%%%%%%%%%%%%%%%%%

Classical action for the bosonic string in Polyakov's formulation
is invariant under 2d reparametrizations and local Weyl rescalings.
As a consequence, all tree components of the metric $g_{\a\b}$ can be
completely gauged away and gravity is classically a nonphysical field.
Quantization of the theory leads to the appearance of an {\it anomaly},
which means that not all classical symmetries are the symmetries of
the quantum theory. We can use the reparametrization invariance to fix the
gauge in the conformally flat form
$$
g_{\alpha\beta}(\xi )=e^{\vphi (\xi )}\eta_{\alpha\beta},       \eqno(1)
$$
where $\eta_{\alpha\beta}=(+,-)$.
The quantum dynamics of the gravitational field in the conformal gauge
is determined by the effective action
$$
W[\vphi ] = -{D-26\over 8\p}I_L \, ,\qquad
I_L[\vphi ] \equiv \int d^2\xi \bigl( {\fr 1 2}\eta^{\a\b}
     \partial_\a\vphi\partial_\b\vphi - \mu^2 e^\vphi \bigr)\, . \eqno(2)
$$
In $D\not=26$ (noncritical string) this dynamics is nontrivial and is
known as the {\it Liouville theory} [1,2].

It is interesting to study the behaviour of this theory under a subgroup of
2d repa\-ra\-me\-trizations, the group of {\it conformal} reparametrizations:
$$
\d_0 g_{\a\b}= (\nabla\cdot\ve )g_{\a\b} \, .                   \eqno(3a)
$$
By using the conformal gauge one finds
$$
\d_0\vphi = \ve\cdot\partial\vphi +\partial\cdot\ve \, ,        \eqno(3b)
$$
with $\ve^\pm=\ve^\pm(\xi^\pm)$. The Liouville action is easily seen to be
{\it invariant} under these transformations.

On the other hand, in the standard Hamiltonian approach the Liouville
action in the conformal gauge {\it is not degenerate\/.} The nondegeneracy
of the action (2) is a natural consequence of the gauge fixing procedure
and implies {\it the absence of first class constraints\/.} As a
consequence, the Hamiltonian origin of the conformal symmetry of
$I_L[\vphi]$ remains a bit obscure. The situation is very simmilar to the
case of the $SL(2,R)$ symmetry in the light--cone gauge [3,4,5].

The objective of the present paper is to study the Hamiltonian
structure of the conformal symmetry in the Liouville theory. It is well
known that gauge symmetries in the Hamiltonian framework are related to
the presence of arbitrary multipliers in the total Hamiltonian [6]. To
clarify the real meaning of this assertion let us consider a dynamical
evolution of a system described by a phase-space trajectory starting
from a given point at time $t=0$. For different choices of arbitrary
multipliers we can solve the Hamiltonian equations of motion and obtain
different trajectories, all starting from the same point and describing
the same physical state. At any time $t>0$ we can pass from one
trajectory to another, without changing the physical state. This
unphysical transition between trajectories at a given time $t$ is
called the gauge transformation. It is clear that the Hamiltonian
definition of gauge symmetries is based on a {\it definite choice of
time\/.} The absence of gauge symmetries in a given Hamiltonian
formalism based on one specific choice of time does not mean that these
symmetries are absent for any other choice. We shall show how the
conformal symmetry of the Liouville theory can be detected by using the
light--cone time variable $\xi^+$.

The conformal transformation  with parameter $\ve^+(\xi^+)$ contains
the inhomogenious term $\partial_+\ve^+$. The presence of this term
suggests that the boundary conditions of $\vphi$ should involve an
arbitrary function of time $\xi^+$, in order to be invariant under the
symmetry transformation. A detailed investigation of the Liouville
theory shows that this is indeed what happens.  Standard Castellani's
method [7] is not general enough to treat gauge symmetries with such
unusual boundary conditions.  This motivated us to develop a general
method for constructing the gauge generators, so that boundary
conditions present in the case of conformal symmetry can also be
consistently described. As a result, the Hamiltonian origin and
structure of the conformal symmetry in the Liouville theory becomes
much more clear.
                                        %%%%%%%%%%%%%%%%%%%%%%%%%%%%%%%%
\section{2. Hamiltonian and constraints in the light--front formalism}

There are several reasons to study relativistic field theories at fixed
light--cone time [8]. Here, the Hamiltonian light--front formalism is used
to clarify the nature of the conformal symmetry in the Liouville theory.

In the light--cone coordinates $\xi^\pm=(\xi^0\pm \xi^1)/\sqrt{2}$
the Liouville action (2) takes the form
$$
I_L=\int d^2\xi \left(
     \partial_-\vphi\partial_+\vphi -\m^2 e^\vphi \right) \, .  \eqno(4)
$$
If we choose $\xi^{+}$ as the time variable, the action becomes
{\it degenerate\/}. The definition of the momentum $\p_\vphi$ leads to the
following primary constraint:
$$
\phi \equiv \p_\vphi -\partial_-\vphi \approx 0 \, .            \eqno(5)
$$

The canonical Hamiltonian is $H_c= \m^2 \int d\xi^- e^\vphi$,
while the total Hamiltonian takes the form
$$
H_T=\int d\xi^- \left( \m^2 e^\vphi + u\phi \right) \, ,        \eqno(6)
$$
where $u$ is, at this stage, an undetermined multiplier.

{C.} The presence of the term $u_0\tilde\phi_0$ in $\tilde H_T$,
with $u_0$ an arbitrary multiplier, means that the dynamics of the
system is characterized by a gauge symmetry. If the gauge
transformation involves a gauge parameter and its time derivative, the
gauge generator has the form
$$
G=\ve G_0 + \dot\ve G_1 \, .
$$
We could now try to construct $G$ by using Castellani's algorithm [7] which
asserts that the phase--space functions $G_0,G_1$ are determined by the
following set of conditions:
$$\eqalign{
G_1 &= C_{PFC} \, , \cr
G_0 + \{ G_1, \tilde H_T\} &= C_{PFC} \, , \cr
\{ G_0,\tilde H_T \} &= C_{PFC} \, , }
$$
where $C_{PFC}$ denotes a primary first class (PFC) constraint,
possibly modified by a surface term. It is natural to start with
$G_1 = \tilde\phi_0$ and calculate $G_0$ from the second equation,
$$
G_0 = {\m^2 \over 2}\int d\xi^- e^{\vphi (\xi )} + \a\tilde\phi_0 \, .
$$
However, the third equation, that represents a kind of consistency test,
fails to be satisfied. Therefore, the  application of this
algorithm leads to {\it contradiction\/.} The resolution of the problem
demands a generalization of Castellani's approach, as will be seen in the
exposition that follows.

The consistency requirements are calculated by using the Poisson
brackets taken at the same time $\xi^+$. By demanding $\{\phi ,H_T\} =0$
one obtains a condition on $u$:
$$
2\partial_- u + \m^2 e^\vphi \approx 0 \, .                     \eqno(7a)
$$
A particular solution of this inhomogenious equation can be chosen
in the form
$$
\hat u = {\m^2\over 4}\int dx
  \e (x -\xi^-) e^{\vphi(x)}\equiv\int dx g(x,\xi^-) \, ,
$$
where the variable $x$ is of the $\xi^-$ type, the dependence on
$\xi^+$ is, for simplicity, not explicitely displayed and
$\e(x)$ is the antisymmetric step function satisfying the
relation $\partial_x\e(x)=2\d(x)$. The quantity $\hat u$ obeys the
antisymmetric boundary conditions:
$$
\hat u_+ = -\hat u _-  \, ,\qquad
\hat u_\pm \equiv \hat u({\xi^-\to \pm\infty}) \, .
$$
General solution for $u$ is obtained by adding an arbitrary function of
$\xi^+$ to $\hat u$:
$$
u(\xi^+,\xi^-) = \hat u(\xi^+,\xi^-) + u_0(\xi^+) \, .          \eqno(7b)
$$
After that the total hamiltonian becomes
$$\eqalign{
&H_T = H' +u_0\phi_0 \, ,\cr
&H'\equiv \int d\xi^- \left( \m^2 e^\vphi +\hat u \phi \right) \, ,
   \qquad \phi_0 \equiv \int d\xi^- \phi \, .}                  \eqno(8)
$$

The Hamiltonian equations of motion are:
$$
\partial_+\vphi = \hat u + u_0 \, ,\qquad
2\partial_+\p =-\m^2 e^\vphi \, .                               \eqno(9)
$$
After differentiating the first equation with respect to $\xi^-$ one
immediately obtains the Lagrangean equation of motion.

The presence of an arbitrary multiplier $(u_0)$ in the total
Hamiltonian is a signal of the existence of a gauge symmetry in the
theory [6]. It is important to note that $u_0$ is a function of only
one coordinate, the time $\xi^+$.

In field theory all kinds of generators, in particular the Hamiltonian
and symmetry generators, are generally {\it nonlocal} functionals of
the phase-space variables. Owing to this, the role of {\it surface
terms} becomes important in establishing the finiteness and
differentiability of the generators, which is needed to properly define
their action [9].

We shall begin the construction of the gauge generators by adopting a
definite {\it asymptotic behaviour} for the basic dynamical variables
$\vphi$ and $\p_\vphi$.

\section{3. Asymptotic bahaviour and surface terms} %%%%%%%%%%%%%%%%%%%

The choice of asymptotics is always guided by some physical requirements.
By demanding the finiteness of the Hamiltonian (finite energy condition)
we easily find that $\exp\varphi$ must decrease
faster than $(\xi^{-})^{-1}$ for large $\xi^{-}$.

The general solution of the Liouville equation is given by
$$
e^\vphi = -{16\over\m^2} {A'(\xi^+)B'(\xi^-) \over
                 \left( 1-A(\xi^+)B(\xi^-)\right)^2 }\, ,
$$
where $A(\xi^+)$ and $B(\xi^-)$ are differentiable functions.
The solution is {\it regular} if the following conditions are satisfied:
$$
\eqalign{ &AB\ne 1 : \cr
          &A'\ne 0,~B'\ne 0: }  \qquad
\eqalign{ &AB<1 \hbox{\rm~~or~~}AB>1\, ,\cr
          &A\hbox{\rm ~and~}B\hbox{\rm ~are~monotonous}\, .}
$$
The second statement assumes that $A$ and $B$ are continuous functions.
 From these relations it follows that $B$ (and $A$) must have at least one
horizontal asymptote.

If $B(\xi^-)$ has two horizontal asymptotes, say $B\sim  \a_\pm +
\b_\pm /(\xi^-)^{s_1}\ (s_1>0)$, then one finds
$$
e^\vphi \sim {a_\pm (\xi^+) \over (\xi^-)^{s_1+1}} \, ,
             \qquad \hbox{\rm when~} \xi^-\to \pm \infty \, ,   \eqno(10a)
$$
where $a_\pm$ are two different functions characterized by the
behaviour of $B(\xi^-)$ at $\xi^-\sim\pm\infty$.
The case when $B(\xi^-)$ has only one horizontal asymptote, say when
$\xi^-\to -\infty$,  while  for $\xi^-\to\ +\infty$ it behaves like
$ (\xi^-)^{s_2}, s_2>0$, leads to
$$
e^\vphi \sim \cases{
b_+/(\xi^-)^{s_2+1} \, ,\quad &\hbox{$\xi^-\to +\infty \, ,$}\cr
b_-/(\xi^-)^{s_1+1} \, ,      &\hbox{$\xi^-\to -\infty \, .$}\cr}
                                                                \eqno(10b)
$$

Thus we see that both cases satisfy the finite energy requirement. To
simplify further exposition we shall consider the case $s_1=s_2=1$, having
in mind that the other cases can be treated in a completly analogous manner.
This leads us to adopt the following asymptotic behaviour for the field
$\varphi$:
$$
\varphi \sim -2\ln\vert\xi^{-}\vert + C_{\pm} + \cO_1 \, ,
        \qquad \xi^-\to \pm\infty \, ,                          \eqno(11)
$$
where $C_{\pm}=C_{\pm}(\xi^{+})$ are two generally different functions,
independent of $\xi^-$, and $\cO_n$ denotes a term that decreases as
$(\xi^-)^{-n}$ or faster when $\xi^-\to\infty$.

To define the asymptotic behaviour of the momentum variable we shall
use the fact that one can demand an arbitrarily fast decrease for those
expressions that vanish on shell, as no solution of the equations of
motion is thereby lost. In accordance with this we define
$$
\pi \sim -{2\over{\xi^{-}}} + \cO_2 \, ,                        \eqno(12)
$$
which ensures the $\cO_2$ behaviour of the constraint $\phi$. It is
now easy to verify that both the Hamiltonian $ H'$ and the constraint
$ \phi_0 $ are well defined, finite quantities.

Let us now check if $\phi_0$ and $ H' $ have well defined functional
derivatives.
A functional $G[q,p]$ has well defined functional derivatives if its
variation can be written as
$$
\d G = \int dx\ [A(x)\d q(x) + B(x)\d p(x) ] \, ,
$$
where $\d q_{,\a}$ and $\d p_{,\a} $ are absent [9].
In general, when the adopted asymptotics does not make surface terms
disappear, this requirement may not be satisfied. For example,
$$
\d\phi_0 = \int d\xi^{-}(\d\pi - \partial_{-}\d\vphi)
         = \int d\xi^{-}\d\pi - \d(C_{+} - C_{-}) \, .
$$
Obviously, $ \phi_0 $ is not a differentiable functional but it can be
improved by adding a suitable surface term. The quantity
$$
\tilde\phi_0 \equiv \phi_0 + C_{+}-C_{-}                          \eqno(13)
$$
has well defined functional derivatives,
$$
{\d\tilde\phi_0 \over \d\varphi(x)} = 0 \, ,\qquad
   {\d\tilde\phi_0 \over \d\pi(x)}  = 1  \, .                      \eqno(14)
$$
Note that its action on local quantities coincides with that of $\phi_0$.

The variation of the Hamiltonian $H'$ has a similar structure:
$$
\d H' = \int dx\,\left\{\left[{\mu^2\over 2}e^{\varphi(x)}
        + \int dy g(x,y)\phi(y)\right]\d\vphi(x)
        + \hat u(x)\d\pi(x) \right\}
        + \hat u_-\d (C_{+} + C_{-})\, .                        \eqno(15)
$$

We see that $H'$ is not differentiable either, but, as opposed to the
case of $\phi_0$, there is no suitable surface term to improve its
differentiability, since $\hat u_-\d (C_+ + C_-)$ can not be put into
the form $\d$(something). Consequently, we are forced to further
specify the asymptotic behaviour of the field $\varphi $ in order to
get rid of the troublesome term. The appropriate restriction of (11) is
given by
$$
C_{+} + C_{-} = 2v(\xi^{+}) \, ,                                \eqno(16)
$$
where $ v(\xi^{+}) $ is an {\it arbitrarily fixed function} of time, so
that $\d(C_++C_-)=0$. Using this condition we easily find:
$$
{\d H'\over\d\varphi(x)} = {\mu^2\over 2}e^{\varphi(x)}
    + \int dy\ g(x,y)\phi(y) \, , \qquad
{\d H'\over\d\pi(x)} = \hat u(x) \, .                           \eqno(17)
$$

Before we definitely adopt the new asymptotics we should check if any
important physical solution is thereby lost. The consistency requirement
on the restriction (16),
$$\eqalign{
\partial_{+}v &= {\fr 1 2}\partial_+ (C_+ + C_- ) \cr
              &={\fr 1 2}[(\partial_{+}\varphi)_{\infty}
                + (\partial_{+}\varphi)_{-\infty}] = u_{0}\, ,} \eqno(18)
$$
shows that $v$ is the arbitrary multiplier and, consequently, it does not
constrain the theory any further. The simpler choice $v=const.$
would obviously be too restrictive, and would completely destroy the gauge
symmetry of the theory.

The asymptotic behaviour of the basic dynamical variables, defined by
(11), (12) and (16), ensures the finiteness and differentiability of the
improved total Hamiltonian
$$
\tilde H_T \equiv H' + u_{0}\tilde\phi_{0} \, .                  \eqno(19)
$$
It has now well defined Poisson brackets with other well defined, nonlocal
quantities.

We now wish to make a few interesting observations.

{A.} The first one is related to the question of how it is seen that the
quantity $\phi_0$, which is a ``linear combination" of the local
constraints $\phi (\xi)$, is also conserved during the time evolution
of the system. The temporal development of $\phi_0$ can not be
calculated through the Poisson bracket $\{ \phi_0,\tilde H_T\}$ since
$\phi_0$ is not a differentiable functional. Instead, we can employ
well defined $\tilde\phi_0$ so that
$$\eqalign{
{d \phi_0 \over d\xi^+} &= {d \tilde\phi_0 \over d\xi^+}
                                      -\partial_+ (C_+ - C_-)  \cr
   &=\{ \tilde\phi_0,\tilde H_T\} -\partial_+ (C_+ - C_-)  \, . }
$$
Using
$$
\{\tilde\phi_0, \tilde H_T\} =\{\tilde\phi_0, H' \}
              \approx -{\m^2 \over 2}\int d\xi^- e^{\vphi (\xi )}  \, ,
$$
one finds that the time consrvation of $\phi_0$  is equivalent to the
condition
$$
\partial_+(C_+ - C_-) +{\m^2\over 2}\int d\xi^- e^{\vphi (\xi )}
                                                       \approx 0 \, .
$$
On the other hand, the Hamiltonian equations of motion yield
$$
\partial_+(C_+-C_-)=(\partial_+\vphi)_\infty -(\partial_+\vphi)_\infty
                   = \hat u_+ - \hat u_-  \, ,
$$
so that the above condition is authomatically satisfied. Therefore,
$\partial_+\phi_0 \approx 0$, as expected.

{B.} The second observation is related to the question of energy
conservation. Since $ \{ \tilde H_T,\tilde H_T \} =0$, we have
$$\eqalign{
{d \tilde H_T\over d\xi^+} &= {\partial \tilde H_T\over\partial \xi^+}
  +\left\{ \tilde H_T , \tilde H_T \right\}  \cr
&={\partial H'\over\partial\xi^+}+\dot u_0\tilde\phi_0 \approx
  2u_0\hat u_- +\dot u_0 (C_+ - C_-) \, . }
$$
We see that the energy is conserved only in the gauge $u_0 =0$.
This is a consequence of the fact that the asymptotic behaviour of $H'$
is time dependent and $\tilde \phi_0$ is {\it not} a constraint due to
the presence of surface terms. As a consequence, the explicit time
dependence of $\tilde H_T$ is absent only in the gauge $u_0=0$.

\section{4. Construction of gauge generators} %%%%%%%%%%%%%%%%%%%%%%%%%

Let us observe that Castellani's algorithm for the construction of
gauge generators {\it is not general enough} to treat the cases in
which the Hamiltonian may nontrivially depend on the arbitrary
parameters of the theory. This is exactly the case with our $\tilde
H_T$ which explicitely depends not only on the multiplier $\dot v$,
but also on $v$:
$$
\tilde H_T = H' + \dot v \tilde\phi_0 \, , \qquad
{\partial H'\over\partial v} = 2 \hat u_- \, ,                  \eqno(20)
$$
as the relations (15), (16) and (18) show. Thus, we are led to generalize
Castellani's method to include a wider class of theories.

\subsub{Generalized conditions for symmetry generators.} %%%%%%%%%%%%%%%
Let us consider a system with finite number of degrees of freedom whose
Hamiltonian explicitely depends on an arbitrary parameter $v(t)$, as
well as on its time derivative $\dot v$:
$$
H_T = H_T (q,p;v,\dot v) \, .                                   \eqno(21)
$$
If there exists a gauge symmetry of the equations of motion involving
only gauge parameter $\ve (t)$ and its time derivative $\dot\ve (t)$,
we assume that the corresponding gauge generator has the form
$$
G(\ve ) \equiv  \ve G_0 + \dot\ve G_1  \, ,\qquad
        G_a = G_a(q,p;v,\dot v) \quad (a=0,1) \, ,              \eqno(22)
$$
so that
$$
\d q =\{ q, G[\ve]\}\, ,\qquad \d p =\{ p, G[\ve]\} \, .
$$
What conditions the functions $G_a$ should satisfy in order that $G[\ve]$
represents a symmetry generator of the theory for arbitrary $\ve (t)$ ?
Obviously, one must demand that the transformed trajectories
$q(t)+\d q(t),\ p(t)+\d p(t)$ also satisfy the Hamiltonian equations of
motion with possibly different parameter $v(t)+\d v(t)$. Thus, the equations
$$
{d\over dt}(\d q)= \d \{ q,H_T \} \, ,\qquad
{d\over dt}(\d p)= \d \{ p,H_T \} \,                            \eqno(23)
$$
must have a solution for $\d v(t)$.
The calculation of the left-hand side of the first equation yields
$$\eqalign{
L ={d\over dt} \{ q, G(\ve ) \} = \ddot\ve\{ q,G_1\}
 &+\dot\ve \big[ \{ \{ q,H_T \},G_1 \}
       + \{ q,G_0+\{ G_1,H_T\}+{\partial G_1 /\partial t} \} \big] \cr
 &+\ve \big[ \{ \{ q,H_T\} ,G_0 \} + \{ q,\{ G_0,H_T\} \}
       +\{ q,{\partial G_0 /\partial t}\} \big] \, ,}
$$
while the calculation of the right-hand side leads to
$$
R= \big[ \dot\ve \{ \{ q, H_T\}, G_1 \}
          +\ve \{ \{ q, H_T\}, G_0 \} \big]
  + \{ q,{\partial H_T / \partial v} \} \d v
  + \{ q, {\partial H_T / \partial \dot v} \} \d\dot v \, .
$$
Since the explicit time dependence of the generators is given only through
the parameters $v$ and $\dot v$, we have
$$
{\partial G_a\over\partial t}={\partial G_a \over\partial v}\dot v
 +{\partial G_a \over\partial\dot v}\ddot v \, .
$$
Similar results are obtained for $p$, too. Combining these relations
the requirements (23) can be written in the form:
$$\eqalign{
&\ddot\ve G_1
      +\dot\ve\big[ G_0+\{ G_1,H_T\}+{\partial G_1 /\partial t}\big] \cr
& +\ve \bigl[ \{ G_0,H_T\} +{\partial G_0 /\partial t} \bigr]
      = {\partial H_T\over\partial v}\d v +
      {\partial H_T\over\partial \dot v}\d\dot v \, .}          \eqno(24)
$$
Here, the equality means an equaity up to quantities that act
trivially on $q$ and $p$, i.e. whose Poisson brackets with $q$ and $p$
weakly vanish. The equation represents a condition for $\d v$
and must hold for every $\ve (t)$ and $v(t)$.
Consequently, it implies:
$$\eqalign{
  G_1 &= \a_1 {\partial H_T\over\partial v} +
         \b_1 {\partial H_T\over\partial\dot v} \, , \cr
  G_0 + \{ G_1,H_T\} + {\partial G_1\over \partial t}
      &= \a_0 {\partial H_T\over\partial v} +
              \b_0 {\partial H_T\over\partial\dot v} \, ,\cr
  \{ G_0,H_T\} +{\partial G_0\over \partial t}
      &= \a {\partial H_T\over\partial v} +
         \b {\partial H_T\over\partial\dot v} \, . }            \eqno(25)
$$

If the total Hamiltonian depends on $\dot v$ as in Eq.(20) but $H'$
does not depend on $v$, the above conditions reduce to those of
Castellani. The dependence of $H_T$ on {\it both} $\dot v$ {\it and} $v$
is the property which demands the generalization of Castellani's method.
It is now clear why the naive application of Castellani's conditions to
the case of conformal symmetry of the Liouville theory did not lead to
the correct answer.

\subsub{Solution of the symmetry conditions.} %%%%%%%%%%%%%%%%%%%%%%%%%
After using the result (25) for $G_a$, the left--hand side of Eq.(24)
takes the form
$$
(\ddot\ve\a_1 +\dot\ve\a_0 +\ve\a){\partial H_T\over \partial v} +
(\ddot\ve\b_1 +\dot\ve\b_0 +\ve\b){\partial H_T\over \partial\dot v} \, .
$$
Assuming that {\it neither} ${\partial H_T/\partial v}$ {\it nor}
${\partial H_T /\partial\dot v}$ vanish, we can solve Eq.(24)
to obtain:
$$\eqalign{
\d v &= \ddot\ve\a_1 + \dot\ve\a_0 + \ve\a  \, ,\cr
\d\dot v &=\ddot\ve\b_1 + \dot\ve\b_0 + \ve\b \, . }
$$
These equations are not consistent unless we require
$$
\ddot\ve\b_1 + \dot\ve\b_0 + \ve\b =
       {d\over dt} (\ddot\ve\a_1 + \dot\ve\a_0 + \ve\a ) \, ,
$$
wherefrom one easily finds the following relations among the
coefficients:
$$
\a_1 = 0 \, ,\qquad \b_1 = \a_0 \, ,\qquad
\b_0 = \dot\a_0 +\a \, ,\qquad \b = \dot\a \, .
$$
We see that there are only two out of six parameters in (25) which
remain undetermined. The generalized conditions for the existence
of gauge generators are therefore found to be:
$$\eqalign{
G_1 &=\a_0 {\partial H_T\over\partial\dot v} \, ,\cr
G_0 + \{ G_1,H_T\} +{\partial G_1\over \partial t} &=
       \a_0 {\partial H_T\over\partial v}
       + (\dot\a_0 + \a) {\partial H_T\over\partial\dot v} \, , \cr
\{ G_0,H_T\} +{\partial G_0\over \partial t} &=
       \a {\partial H_T\over\partial v}
        + \dot\a{\partial H_T\over\partial\dot v} \, . }        \eqno(26)
$$
Note that this holds only when $H_T$ nontrivialy depends on both
$v$ and $\dot v$. If one of the quantities
${\partial H_T /\partial v}$ or ${\partial H_T /\partial\dot v}$
vanishes, the conditions (26) become much simpler, and boil down
to Castellani's conditions if $v$ (or $\dot v$) stands for the usual
arbitrary multiplier of the theory. In the case when both
${\partial H_T /\partial v} $ and
${\partial H_T /\partial\dot v}$ vanish the theory does not possess
any gauge symmetry at all (in this case there are no arbitrary
parameters in $H_T$).

\section{5. Conformal symmetry in Liouville theory} %%%%%%%%%%%%%%%%%%%%

Let us, now, apply the new method to the Liouville Hamiltonian (20),
which is obtained in the light--front formalism with time $\t=\xi^+$.
As $\tilde\phi_0$ does not depend on either $v$ or $\dot v$ and $H'$
depends on $v$ alone, the conditions (26) simplify to
$$\eqalign{
G_1 &=\a_0 \tilde\phi_0 \, ,\cr
G_0 + \{ G_1,\tilde H_T\} +{\partial G_1\over \partial\t} &=
       \a_0 {\partial H'\over\partial v}
       + (\dot\a_0 + \a) \tilde\phi_0 \, , \cr
\{ G_0, \tilde H_T\} +{\partial G_0\over \partial\t} &=
       \a {\partial H'\over\partial v}
        + \dot\a\tilde\phi_0  \, . }                            \eqno(27)
$$
The first two equations are easily solved to give
$$
G_1 = \a_0 \tilde\phi_0 \, ,\qquad  G_0
    = \a_0 \left( {\partial H'\over\partial v} - \left\{
      \tilde\phi_0,H'\right\} \right) + \a\tilde\phi_0 \, ,
$$
where the parameters $\a $ and $\a_0 $ should be determined, if
possible, from the third requirement. Using Eqs.(14), (17) and (20)
we find
$$
G_1 = \a_0 \tilde\phi_0 \, ,\qquad
G_0 = \a_0 H' + \a\tilde\phi_0 \, ,
$$
with
$$
\{ G_0,\tilde H_T \} +{\partial G_0\over \partial\t} =
     \a {\partial H'\over\partial v} + \dot\a\tilde\phi_0
                    + (\dot\a_0 + \a_0 \dot v - \a )H' \, .
$$
Thus, the third requirement in (27) will be satisfied if
$$
\dot\a_0 + \a_0 \dot v - \a = 0 \, ,
$$
the simplest solution of which is
$$
\a_0 = 1 \, ,\qquad \a = \dot v \, .
$$
These values determine the final form of the gauge generators:
$$
G_1 = \tilde\phi_0 \, ,\qquad  G_0 = \tilde H_T \, .            \eqno(28a)
$$
It is easily checked that the generator
$$
G[\ve]\equiv\int d\xi^-\left[ \ve \tilde H_T +
                       (\partial_+\ve)\tilde\phi_0 \right]       \eqno(28b)
$$
indeed produces the conformal gauge transformations:
$$\eqalign{
& \d \vphi =\{\vphi,G[\ve]\}=\ve\partial_+\vphi +\partial_+\ve \, , \cr
& \d\p =\{\p ,G[\ve]\} = \ve\partial_+\p \, .  }                 \eqno(29)
$$

\vskip.3cm
In conclusion, we studied the conformal symmetry of the Liouville
theory in the Hamiltonian formalism by going over to the light--front
time. The corresponding gauge generators were found after ganeralizing
the existing methods, so that the dependence of $H_T$ on both $v$
and $\dot v$, stemming from the specific boundary conditions, could be
consistently taken into account.

\vfill\eject

\section{References:}

\item{1.} A. M. Polyakov, Phys. Lett. {\bf B103} (1981) 207.
\item{2.} E. D'Hoker and D. H. Phong, Rev. Mod. Phys.
{\bf 60} (1988) 917.
\item{3.} A. M. Polyakov, Mod. Phys. Lett. {\bf A2} (1987) 893;
% V. G. Knizhnik, A. M. Polyakov and A. B. Zamolodchikov,
% Mod. Phys. Lett. {\bf A3} (1988) 819.
\item{4.} Ed. Sh. Egorian and R. P. Manvelian, Mod. Phys. Lett.
{\bf A5} (1990) 2371;
J. Barcelos-Neto, Constraints and hidden symmetry in 2D-gravity,
Univ. Rio de Janeiro preprint IF/UFRJ/92/21 (1992);
\item{5.} M. Blagojevi\'c, M. Vasili\'c and T. Vuka\v sinac,
Hamiltonian analysis of $SL(2,R)$ symmetry in Liouville theory,
Institute of Physics preprint IF-S12/93 (1993).
\item{6.} P. A. M. Dirac, Rev. Mod. Phys. {\bf 21} (1948) 392;
K. G. Wilson, Nucl. Phys. {\bf B} (Proc. Suppl.) {\bf 17}
(1990);  Prem. P. Srivastava, Constraints and Hamiltonian in
Light-Front Quantized Field Theory, preprint DFPF/9/TH/58 (1992).
\item{7.} P. A. M. Dirac, {\it Lectures on quantum mechanics\/,}
Yeshiva University --- Belfer Graduate School of Science (Academic,
New York, 1964); A. Hanson, T. Regge and C. Teitelboim,
{\it Constrained Hamiltonian Dynamics\/,} (Academia Nationale del
Lincei, Rome, 1976);
K. Sundermeyer, {\it Constrained Dynamics\/,} (Springer, Berlin, 1982).
\item{8.} T. Regge and C. Teitelboim, Ann. of Phys. (NY) {\bf 88}
(1974) 286; R. Benguria, P. Cordero and C. Teitelboim, Nucl. Phys.
{\bf B122} (1977) 61; P. J. Steinhardt, Ann. of Phys. (NY) {\bf 128}
(1980) 425.
\item{9.} L. Castellani, Ann. Phys. (N. Y.) {\bf 143} (1982) 357; see
also X. Gr\`acia and J. M. Pons, Ann. of Phys. {\bf 187} (1988) 355.

\bye